\documentclass[runningheads]{llncs}

\usepackage{hyperref}
\usepackage{multicol}
\usepackage[final]{changes}
\usepackage{url}
\usepackage{amsmath}
\usepackage{bm}
\usepackage{multirow}
\usepackage{tcolorbox}
\usepackage{rotating}
\usepackage{latexsym,amssymb,amsmath}
\usepackage{makecell}
\usepackage{xspace}
\usepackage{paralist}
\usepackage{wrapfig}
\usepackage{adjustbox}
\usepackage{cite}

\newcolumntype{C}{>{\centering\arraybackslash}X}

\newcounter{dummy} 
\newcounter{dummy1} 
\newcounter{dummy2}
\newcounter{dummy3} 
\newcounter{dummy4}
\newcounter{dummy5} 
\newcounter{dummy6}

\usepackage[thmmarks,amsmath]{ntheorem}

\theoremseparator{.}
\theorembodyfont{\itshape}
\theoremsymbol{$\triangleleft$}
\newtheorem{theorem}[dummy]{Theorem}
\newtheorem{definition}[dummy2]{Definition}
\newtheorem{proposition}[dummy3]{Proposition}

\usepackage{booktabs}
\usepackage{apptools}
\usepackage{chngcntr}
\newtheorem{lemma}[dummy1]{Lemma}

\theorembodyfont{\normalfont}
\newtheorem{example}[dummy4]{Example}
\newtheorem{remark}[dummy5]{Remark}

\theoremstyle{nonumberplain}
\theoremheaderfont{\itshape}
\theorembodyfont{\normalfont}

\theoremseparator{.}
\theoremsymbol{\ensuremath{\dashv}}
\newtheorem{proof}[dummy6]{Proof}

\qedsymbol{\ensuremath{\dashv}}

\usepackage{todonotes}

\usepackage{graphicx}
\usepackage{xcolor,color}
\usepackage{subfig}
\usepackage{calc}

\usepackage{tabularx}
\usepackage{booktabs}

\usepackage{ulem}

\usepackage{kbordermatrix}
\usepackage{amsmath,amsfonts}
\usepackage{braket}
\usepackage{xfrac}

\usepackage{pifont}
\usepackage{amssymb}
\usepackage{paralist}
\usepackage[inline]{enumitem}

\newcommand{\const}[1]{\mathsf{#1}}

\newcommand{\DeclModel}{\mathcal{D}}
\newcommand{\Activities}{A}

\newcommand{\approptoinn}[2]{\mathrel{\vcenter{
			\offinterlineskip\halign{\hfil$##$\cr
				#1\propto\cr\noalign{\kern2pt}#1\sim\cr\noalign{\kern-2pt}}}}}

\def\EqualityHolds{itholds}
\ifdefined\EqualityHolds

\else

\fi

\usepackage{stmaryrd}
\newcommand{\sem}[1]{\ensuremath{\llbracket #1 \rrbracket}}

\newcommand{\Next}{{\ensuremath\raisebox{0.25ex}{\text{\scriptsize$\bigcirc$}}}}
\newcommand{\Until}{\ensuremath{\mathbin{\mathcal{U}}}}

\allowdisplaybreaks

\newif\iffull\fulltrue

\tikzstyle{pconstraint} = [
draw=lightBlue!80,
]

\tikzstyle{pwords} = [
text=lightBlue!80
]

\newcommand{\translation}[2]{\ensuremath{\uptau_{#1}^{#2}}}

\usepackage{graphicx}
\usepackage{actuarialsymbol2}
\usepackage{upgreek}

\usepackage{linegoal}
\usepackage{changepage}
\newlength{\previousLine}

\begin{document}

\title{On Declare MAX-SAT and a finite\\  Herbrand Base for data-aware logs}
\author{Giacomo Bergami}
\institute{\texttt{bergamigiacomo@gmail.com}}
%
%

%

%
%
\maketitle              
\linespread{0.92}
\vspace{-0.5cm}

\begin{abstract}
This technical report provides some lightweight introduction motivating the definition of an alignment of log traces against Data-Aware Declare Models potentially containing correlation conditions. This technical report is only providing the intuition of the logical framework as a feasibility study for a future formalization and experiment section. 
\end{abstract}

\section{Working assumptions}

In this draft paper\footnote{\texttt{https://github.com/jackbergus/bpm21-dataalign/commits/}\\\texttt{fbd770f998e1c9f439db31e1773b31ef01533b5a}}, we are considering the  alignment problem for log traces $\sigma\in\mathcal{L}$ with respect to Declare Data-Aware models $\mathcal{M}$, where traces might have data payloads. We are interested in reformulating the alignment problem as a satisfiability of such a model over a single world, e.g., each single log trace. As a common assumption in current literature, each log trace has a finite length.  As a consequence, we are working under a Closed World Assumption, as each time we pick a trace $\sigma$ and test it against the logical model $\mathcal{M}$; this approach can be also trivially generalised to a whole log $\mathcal{L}$ after imposing that the log set contains a finite number of log traces.

We assume that each log trace represents the ``sequentialisation'' of possibly concurrent processes' activities, thus implying that we can always provide a complete total order of the traces' events by exploiting their timestamp information. Furthermore, it will be always possible to determine a bijection associating the timestamp to the position of an event within the ``sequentialised'' trace.

In particular, we want to show that such problem  
%
can be formulated as a MAX-SAT problem: the perfect alignment (zero-cost) will correspond to a 1 returned by the MAX-SAT problem. This is possible since, in current literature, each Declare Data Aware clause is defined independently from the other clauses in the model, and therefore we can test the satisfiability of each clause $C_i\in\mathcal{M}$ at a time: therefore a log trace satisfies the model iff. the log trace satisfies each clause belonging to such model, $\forall \sigma, \mathcal{M}.\; \sigma\vDash\mathcal{M}\Leftrightarrow \left(\forall C_i\in\mathcal{M}. \sigma\vDash C_i\right)$. Please observe that this is a completely different assumption from recent SAT-solvers \cite{LiPZVR20}, where the satisfiability of a model $\mathcal{M}$ is defined over the set of all the possible model traces, thus making the satisfiability problem semi-decidable; furthermore, the latter approach cannot be trivially exploited for solving a trace alignment problem as we intend.
As MAX-SAT is generally an NP-Complete problem, finite models enable tractable solutions. In particular, we will show that it is always possible to define a decidable algorithm in our sketched scenario.

\section{FOL formulation}
(Data) \emph{payloads} are finite functions $p\in V^K$, where $K$ is a finite set of keys and $V$ is a (finite) set of values. We distinguish the trace keys ($K_t$) from the event keys ($K_e$), such that $K=K_t\cup K_e$ with $K_t\cap K_e=\emptyset$. We denote $\bot$ as an element $\bot\notin V$: we can now denote $p(k)=\bot$ for $k\notin\textup{dom}(p)$.  An \emph{event} $e_j$ is a pair $\Braket{\const{a},p}\in \Activities\times V^K$, where $A$ is a finite set of activity labels and $p$ is a finite function describing the data payload. A \emph{trace} $\sigma$ is a ordered sequence of distinct events $e_1,\dots,e_n$ for which each event associates the same values to the same trace keys ($\forall \Braket{\const{a}_i,p_i},\Braket{\const{a}_j,p_j}\in \sigma.\forall k\in K_t. p_i(k)=p_j(k)$). A \emph{log} $\mathcal{L}$ is a finite set of {traces}, where each trace $\sigma$ is uniquely associated by a numeric case-id, $\textup{case}(\sigma)\in V$ . Last, we can freely assume that there exists a specific timestamp event key $T\in K_e$, such that for each trace $\Braket{\const{a}_1,p_1},\dots,\Braket{\const{a}_i,p_i},\dots,\Braket{\const{a}_n,p_n}$ it always exists
 a bijection $i\overset{t}{\leftrightarrow} p_i(T)$, thus implying that temporal aspects of the events can be directly represented as payload values.

Values can represent either categorical data ($U^*$), numerical data ($F_{(\beta,t,\lambda,\omega)}$),
or hierarchical data ($H$). Given $U$ the set of the UTF-32 characters, $U^*$ represents the set of all the possible strings, for which  a lexicographical ordering $\preceq_{U^*}$  exists. Numerical data can be represented via a finite number system $F_{(\beta,t,\lambda,\omega)}=\{\alpha\in\mathbb{R}|\alpha=\pm (\sum_{i=0}^{t-1}\alpha_i\beta^i)\beta^p,\;\lambda\leq p\leq \omega\}\cup\Set{0}$, where IEEE floats are represented by $F_{(2,23,-126,127)}$, for which it trivially exists an ordering $\preceq_{\mathbb{R}}$. Hierarchical data can be described as a partially ordered set $(H,\preceq_H)$, where $\preceq_H$ determines an is-a relationship among entities. Please note that is always possible to compose multiple hierarchies into one single hierarchy via graph cartesian product \cite{BergamiBM20}. 

At this point, we want to prove that each log $\mathcal{L}$ can be described as a finite model: 
%
%
this is a sufficient requirement to make any FOL formula decidable \cite{harrop},
%
as we just need to enumerate larger and larger finite interpretations of a given trace until we find one in which it holds.
\begin{lemma}\label{lem:hu}
	Values can be represented as a finite partially ordered set $\mathcal{V}=(V\cup\{\top,\bot\},\preceq_V)$.
\end{lemma}
\begin{proof}
	Values in $V$ can be described by either categorical data, numerical data, or hierarchical data. For categorical data in $U^*$, it always exists a trivial lexicographical ordering $\preceq_{U^*}$. Given that finite number systems represent a finite set of real numbers in $\mathbb{R}$, numerical data expressed as such always admits the same ordering as the real numbers, which are a poset $(\mathbb{R},\preceq_{\mathbb{R}})$. Last, given that hierarchical data can be expressed as DAGs $(H,\preceq_H)$, where $H$ are the hierarchy's entities and $\preceq_H$ expresses the is-a relationships, $(H,\preceq_H)$ is a poset \cite{BergamiBM20}. Given that $U^*$, $H$, and $F_{(2,23,-126,127)}$ are sets of distinct elements, we can represent $V$ as a finite set of values completely describing the payloads of a log $\mathcal{L}$ as follows:
	\[V=\bigcup_{e_i\in \mathcal{L}}\Set{p(K)|\Braket{\const{a}, p}\in e_i}\backslash\{\bot\}\]
	After defining $\bot$ as the minimal element of $V$ and $\top$ the maximal element of such set, we can define an ordering $\preceq_V$ as follows:
	\begin{align*}
	\forall u,v\in V.\; &u\preceq_V v\Leftrightarrow &&\;(u\in U^*\wedge v\in U^*\wedge u\preceq_{U^*}v)\\
	& &&\vee (u\in F_{(\beta,t,\lambda,\omega)}\wedge v\in F_{(\beta,t,\lambda,\omega)}\wedge u\preceq_{\mathbb{R}}v)\\
	& &&\vee (u\in H v\in H\wedge u\preceq_{H}v)\\
\forall s\in\min(U^*\cap V).\;&\bot\preceq_V  s\\
	\forall s\in\max(U^*\cap V).\;&s\preceq_V  \top\\
	\forall f\in\min(F_{(2,23,-126,127)}\cap V).\; &\bot\preceq_V f \\
	\forall f\in\max(F_{(2,23,-126,127)}\cap V).\; &f\preceq_V \top \\
	 \forall h\in \min(H\cap V).\;&\bot\preceq_V  h\\
	 \forall h\in \max(H\cap V).\;&h\preceq_V  \top\\
	\end{align*}
\end{proof}

\begin{lemma}
	Each log $\mathcal{L}$ can be represented as a finite model $\Delta(\mathcal{L})$.
\end{lemma}
\begin{proof}
	Let  $A\cup\{\mathtt{``}{\preceq_V}\mathtt{"}\}$ be the set of the predicate symbols. Each predicate symbol $\const{a}\in A$ represents an $(|K|+1)$-ary predicate, while $\mathtt{``}{\preceq_V}\mathtt{"}$ represents a binary predicate. Given that Lemma \ref{lem:hu} completely characterizes the Herbrand's Universe $U(\mathcal{L})=\mathcal{S}$ for the finite log $\mathcal{L}$ as a finite set and given that the set of symbols is also finite, the resulting Herbrand's Base $B(\mathcal{L})$ for the log $\mathcal{L}$ is also a finite set of grounded atoms. Given that any possible Herbrand model resulting from such base is finite, then the set of all the possible {worlds} is finite, and each {world} is finite. Therefore, the resulting model $\Delta(\mathcal{L})$ is finite.

	Each $i$-th event $\Braket{\const{a}_i,p_i}$ from a trace $\sigma\in\mathcal{L}$ can be represented for $K=\Set{T,k_2,\dots,k_n}$ as a $(n+1)$-ary predicate $\const{a}_i(\textup{case}(\sigma),i,p_i(k_2),\dots,p_i(k_n))$. Therefore, $\mathcal{L}$ can be represented as follows:
	\begin{align*}
	\Delta(\mathcal{L})=&\Set{\mathtt{``}{\preceq_V}\mathtt{"}(u,v)|u,v\in B(\mathcal{L})\wedge u\preceq_V v}\cup\\
	&\Set{\const{a}_i(\textup{case}(\sigma),i,p_i(k_2),\dots,p_i(k_n))|\Braket{\const{a}_i,p_i}\in \sigma,\sigma\in\mathcal{L}}
	\end{align*}
\end{proof}

Walking in the footsteps of \cite{GiacomoMM14}, we can provide a semantics to the LTL$_f$ formulae over the previously given Herbrand Base $\Delta(\mathcal{L})$. We can {interpret} an LTL$_f$ formula {in negation normal form} as $\textit{fol}(\phi,\textup{case}(\sigma))$ for each trace $\sigma$
as follows:
\[\textit{fol}(\phi,c)=\begin{cases}
\exists x_1,\dots,x_m\in \mathcal{S}. \textit{pl}(\phi,c,{\color{magenta}1}) & \Set{x_1,\dots,x_m}=\textit{FV}(\textit{pl}(\phi,c,{\color{magenta}1}))\\
\textit{pl}(\phi,c,{\color{magenta}1}) & \textup{oth.}\\
\end{cases}\]
{where \textit{pl} is inductively defined over $\phi$ as follows:}
\[\textit{pl}(\phi,c,{\color{magenta}t})=\begin{cases}\top & \phi=\top\\
\const{a}(c,{\color{magenta}t},\actsymb{({x_{\const{a}\mathbf{P}}})}{\color{black}2}[\color{magenta}t],\dots,\actsymb{({x_{\const{a}\mathbf{P}}})}{n}[\color{magenta}t])\wedge \mathbf{P}  &\phi=\const{a}\wedge \mathbf{P}\\

\neg \textit{pl}(\phi,c,{\color{magenta}t}) & \phi=\neg \phi\\
\textit{pl}(\phi,c,{\color{magenta}t+1}) & \phi=\Next \phi\\	 
\textit{pl}(\phi_1,c,{\color{magenta}t})\wedge \textit{pl}(\phi_2,c,{\color{magenta}t}) & \phi= \phi_1\wedge \phi_2\\
\bigvee_{\color{magenta}t\leq\tau\leq|\sigma|}\textit{pl}(\phi_2,c,{\color{magenta}\tau})\wedge\bigwedge_{\color{magenta}t\leq u<\tau}\textit{pl}(\phi_1,c,{\color{magenta}u}) & \phi= \phi_1\Until \phi_2\\

\end{cases}\] 
where each predicate $\mathbf{P}$ is a relation predicate that can be expressed in terms of $\mathtt{``}\preceq_V\mathtt{"}$. Please observe that, differently from \cite{DBLP:conf/bpm/MaggiDGM13} the existential quantification over the payload values was pulled at the beginning of the formula as in the tuple relational calculus \cite{10.1145/362384.362685}  to guarantee the expression of join conditions:
\begin{itemize}
	\item given a constant value $\kappa$, the condition $\textit{Cond}:=\const{a}_i.K_j\preceq_V \kappa$ to be tested at time $t$ can be expressed as $\mathbf{P}:=(x_{\const{a}_i\textit{Cond}})_j^t\preceq_V\kappa$
	\item the join condition $\textit{Join}:=\const{a}_i.K_j\preceq_V \const{a}_h.K_k$ where $\const{a}_i$ happening at time $t$ {temporally} follows $\const{a}_j$  can be expressed as $\mathbf{P}:=\bigvee_{\tau<t,\textit{Cond}}(x_{\const{a}_i\textit{Join}})_j^t\preceq_V (x_{\const{a}_h\textit{Cond}})_k^\tau$
\end{itemize}

{By the working assumptions, a Declare Data Aware model $\mathcal{M}$ is a set of instantiated Declare Data Aware templates $\mathcal{M}=\Set{C_i}_{i\leq n}$, where each template $C_i$ can be expressed in LTL$_f$ via a translation function $\translation{\textup{D}}{\textup{LTL}_f}$, such that $\translation{\textup{D}}{\textup{LTL}_f}(C_i)$ is a well-formed LTL$_f$ formula. The usual semantic interpretation of such model $\mathcal{M}$ is the conjunction of all of such interpretations, and therefore $\sem{\mathcal{M}}=\bigwedge_{C_i\in\mathcal{M}}\translation{\textup{D}}{\textup{LTL}_f}(C_i)$. Therefore, the satisfiability of a log trace $\sigma\in\mathcal{L}$ can be expressed as $\sigma\vDash\mathcal{M}\Leftrightarrow \bigwedge_{C_i\in\mathcal{M}}\textit{fol}\left(\translation{\textup{D}}{\textup{LTL}_f}(C_i),\;\textup{case}(\sigma)\right)$.}

After observing that the only instance of the existential quantifier is merely a way to bound the payload values to a predicate via variables $({x_{\const{a}\mathbf{P}}})_2^t,\dots,({x_{\const{a}\mathbf{P}}})_k^t$, and given that at each timestamp $t$ only one event can occur in a trace with case-id\footnote{In fact, it is possible to determine a bijection between an event's timestamp within a log trace and its position within such a trace, and each trace within the log has a finite length.} $c$,
then the former FOL fragment is decidable via quantifier elimination\footnote{Given a trace having a specific (and unique)  case-id, it  always exists an unique event happening in a given instant of time, and therefore we can always get its payload.}. 
In particular, the former definition implies that the it is always possible to test whether a trace $\sigma$ is a {possible world} for a given Data Aware Declare template $\phi$ in polynomial time over the size of the trace and in exponential time over the query size, similarly to SQL queries \cite{DBLP:conf/stoc/Vardi82}. 

\subsection{Towards a Knowledge Base representation for MAX-SAT}\label{ref:dddmm}
In some other use cases, the model $\DeclModel$ is unknown, and we want to mine the set of plausible rules from a log. In this case, we are interested in knowing which and how many log traces $\sigma_L$ satisfy (albeit approximately) a Declare constraint $C_i$. This requires to visit the same log dataset multiple times for multiple candidate constraints $C_i\in\mathcal{M}$: although it already exists a previous attempt at minining declarative models via relational databases \cite{SSRSA}, state-of-the-art interpretation of relational queries face inefficient implementation of aggregation operations \cite{BergamiPM18}. It can be showed that some counting constraints (e.g., \texttt{existence}, \texttt{init}) can be efficiently computed while loading the data traces within the relational database thus avoiding the counting cost: however, the authors did not consider this possibility in their implementation. Furthermore, row-oriented systems such as the Microsoft SQL Server exploited in \cite{SSRSA} are not particularly query efficient if compared to column-based storage \cite{IdreosGNMMK12}, where each $n$-ary relation $r(\texttt{id},A_1,\dots,A_n)$ can be decomposed into $n$ relations $r_i(\texttt{id},A_i)$. Last, at the time of the writing, no relational database system is capable of running multiple queries contemporaneously while minimising the data access and visit time to the log space: in this paper, we provide a from-scratch implementation of a in-memory relational database for data-driven declare mining enabling an efficient parallel implementation of the query plan, thus making the mining of Data-Driven Declare Models particularly efficient. In particular, given a log $\mathcal{L}=\Set{\sigma_1,\sigma_2,\sigma_3}$, where $\sigma_1=\mathsf{aaab}$, $\sigma_2=\mathsf{bbbba}$, and $\sigma_3=\mathsf{cbcbc}$, we might represent those traces in these two tables:

\begin{center}
	\begin{tabular}{c|c|c}
	\toprule
	\multicolumn{3}{c}{\texttt{CountTemplate}}\\
	\toprule
	\texttt{act} & $\sigma_{\texttt{id}}$ & \texttt{count}\\
	\midrule
	$\mathsf{a}$& 1 & 3 \\
	$\mathsf{a}$& 2 & 1 \\
	$\mathsf{a}$& 3 & 0 \\
	$\mathsf{b}$& 1 & 1 \\
	$\mathsf{b}$& 2 & 4 \\
	$\mathsf{b}$& 3 & 2 \\
	$\mathsf{c}$& 1 & 0 \\
	$\mathsf{c}$& 2 & 0 \\
	$\mathsf{c}$& 3 & 3 \\
	\bottomrule
\end{tabular} \begin{tabular}{c|c|c|c|c}
\toprule
\multicolumn{5}{c}{\texttt{Act}}\\
\toprule
 \texttt{act} & $\sigma_{\texttt{id}}$ & \texttt{time} & \texttt{next} & \texttt{prev}\\
\midrule
 $\mathsf{a}$ &1 & 0.00 &  2 & $-\infty$\\ 	
 $\mathsf{a}$ &1 & 0.33 &  3 & 1\\ 			
 $\mathsf{a}$ &1 & 0.66 &  \textbf{5} & 2\\ 			
 $\mathsf{a}$ &2 & 1.00 &  $+\infty$ & \textbf{9}\\ 	
 $\mathsf{b}$ &1 & 1.00 &  $+\infty$ & 3\\	
 $\mathsf{b}$ &2 & 0.00 &  \textbf{7} & $-\infty$\\ 	
 $\mathsf{b}$ &2 & 0.25 &  \textbf{8} & \textbf{6}\\ 			
 $\mathsf{b}$ &2 & 0.50 &  \textbf{9} & \textbf{7}\\ 			
 $\mathsf{b}$ &2 & 0.75 &  \textbf{4} & \textbf{8}\\ 			
 $\mathsf{b}$ &3 & 0.25 &  \textbf{13} & \textbf{12}\\ 		
 $\mathsf{b}$ &3 & 0.75 &  14 & \textbf{13}\\ 		
 $\mathsf{c}$ &3 & 0.00 &  \textbf{10} & $-\infty$\\ 
 $\mathsf{c}$ &3 & 0.50 &  \textbf{11} & \textbf{10}\\ 		
 $\mathsf{c}$ &3 & 1.00 &  $+\infty$ & \textbf{11}\\ 
\bottomrule
\end{tabular}\end{center}

When traces contain payloads, we can consider generating a table \texttt{AttributeK}$_i$ for each $\texttt{K}_i\in K$ so to assess  data predicates $\mathbf{P}$ as follows:
\begin{tabular}{c|c|c}
	\toprule
	\multicolumn{3}{c}{\texttt{AttributeK}$_i$\quad($\mathbf{P}$)}\\
	\toprule
	\texttt{act} & \texttt{value} & \texttt{ActOffset}\\
	\midrule
	$\cdots$ & $\cdots$ & $\cdots$\\
	
	\bottomrule
\end{tabular}

In the following expressions, the alignment returns a pair, where the first element is the candidate trace for a model, and the second argument is the alignment similarity. In this draft paper, we are just going to give some preliminary examples, and we leave the others as an exercise to the reader.
\begin{itemize}

\item $\mathcal{A}(\mathsf{init}(A),\mathcal{L})=\pi_{\sigma_{\texttt{id}},\tilde{{t}}}(\textit{Calc}_{{\tilde{{t}}}:=1-\texttt{time}}(\sigma_{\texttt{act}=A}(\texttt{Act}))$
\item $\mathcal{A}(\mathsf{end}(A),\mathcal{L})=\pi_{\sigma_{\texttt{id}},\texttt{time}}(\sigma_{\texttt{act}=A}(\texttt{Act})))$
\item $\mathcal{A}(\mathsf{exactly}(A,n),\mathcal{L})=\pi_{\sigma_{\texttt{id}},\tilde{t}}(\textit{Calc}_{\tilde{t}:=1-\frac{|n-\texttt{count}|}{\mathsf{len}(\sigma_\texttt{id})}}(\sigma_{\texttt{act}=A}(\texttt{CountTemplate})))$
\item $\mathcal{A}(\mathsf{existence}(A,n),\mathcal{L})=\pi_{\sigma_{\texttt{id}},\tilde{t}}(\textit{Calc}_{\tilde{t}:=|\texttt{count}\geq n|\texttt{?}\;1\;\texttt{:}1-\frac{|n-\texttt{count}|}{\mathsf{len}(\sigma_\texttt{id})}}(\sigma_{\texttt{act}=A}(\texttt{CountTemplate})))$
\item \begin{align*}
\mathcal{A}(\mathsf{respexistence}(&A,B,\mathbf{P}),\mathcal{L})\\
&=\mathsf{notexists}^1(A)\oplus\mathsf{ifte}_{\sigma\mapsto(A\wedge \mathbf{P})(\sigma)}(\mathsf{exists}^1(B){\color{red}\oplus\mathsf{notexists}^{c}(B)},\;\top^1)
\end{align*}
\end{itemize} {\color{red}with $c\in[0,1]\subseteq\mathbb{R}_{\geq 0}$}

\begin{align*}
\mathsf{respexistence}(A,B,\mathbf{P})&=\Diamond(A\wedge \mathbf{P})\Rightarrow\Diamond B\\
	&=\sigma\mapsto [\exists t. \lambda(\sigma_t)=A\wedge \mathbf{P}(\sigma_t)]\Rightarrow [\exists t. \lambda(\sigma_t)=B]\\
	&=\sigma\mapsto\texttt{if}\;{|\Set{t\leq |\sigma|\;|\;\lambda(\sigma_t)=A}|=0}\;\texttt{then}\\
	&\qquad\qquad \texttt{return}\;1\\
	&\qquad\quad\;\; \texttt{else if}\; \lambda(\sigma_t)=A\wedge\mathbf{P}(\sigma_t)\;\texttt{then}\\
	&\qquad\qquad \texttt{return\;\{\;if}\;(\exists t.\lambda(\sigma_t)=B)\;   \texttt{then}\;1\;\texttt{else}\;{\color{red}c}\texttt{\}}\\
	&\qquad\quad\;\; \texttt{else return}\;1\\
\end{align*}
\[\mathsf{exists}^i(A)=\Set{\braket{i,l}|l\in\pi_{\sigma_{\texttt{id}}}(\sigma_{\texttt{act}=A\wedge \texttt{count}\neq0}(\texttt{CountTemplate}))}\]
\[\mathbf{P}=\bigwedge_{\texttt{K}_i\theta} \mathbf{P}_{\texttt{K}_i\theta}=\bigcap \sigma_{\mathbf{P}_{\texttt{K}_i\theta}}(\texttt{AttributeK}_i)\]
\[\mathsf{notexists}^i(A)=\Set{\braket{i,l}|l\in\pi_{\sigma_{\texttt{id}}}(\sigma_{\texttt{act}=A\wedge \texttt{count}=0}(\texttt{CountTemplate}))}\]
\noindent
Given the binary disjoint union $\oplus$, if the left and the right operands have no elements in common, the weighted union of two sets $A\oplus B$ is defined as follows:
\[A\oplus B=\Set{\braket{a,p}|\braket{a,p}\in A\veebar \braket{a,p}\in B}\]
The intersection of two weighted set is the following:
\[A\cap B=\Set{\braket{a,pq}|\braket{a,p}\in A \wedge \braket{a,q}\in B}\]
Still, this definition of intersection excludes the elements satisfying either $A$ or $B$, thus only including traces satisfying both constraints. In order to overcome to this limitation, we can provide the following definition of n-ary set intersection:
\[\bigcap_n S_n = \Set{\Braket{a,\frac{1}{n}\sum_{\braket{a,p_i}\in S_i}^{i\leq n}p_i}|\exists j\leq n. \braket{a,\_}\in S_j}\]

\bibliographystyle{splncs04}
\bibliography{main}

\appendix

\counterwithin{lemma}{section}
\renewcommand{\thelemma}{\thesection\arabic{lemma}}


\end{document}